# Simulating Araponga - The High-Resolution Diffractometer of Brazilian Multipurpose Reactor

A. P. S. SOUZA[1a], L. P. de OLIVEIRA[a], F. A. GENEZINI[a], and A. dos SANTOS[a]

[a]Instituto de Pesquisas Energéticas e Nucleares - IPEN, 05508-000, São Paulo, São Paulo, Brazil

[1]alexandre.souza@ipen.br

**ABSTRACT**

The Brazilian Multipurpose Reactor (RMB) is a fundamental project that aims to turn Brazil into a self-sufficient country in the production of radioisotopes and radiopharmaceuticals to supply the Unified Health System (SUS) as much as the private institutions. In addition, the RMB project describes other applications as irradiation and testing of nuclear fuels and structural material analysis, for instance. There are many techniques in the project to study structural aspects of materials, where neutron diffraction represents one of the priorities for implementation. This technique will take place mainly on two diffractometers on Thermal Neutron Guide 1 (TG1), namely Araponga, a high-resolution diffractometer, and Flautim, a high-intensity diffractometer. In this work, we study the performance of the Araponga diffractometer through McStas simulations with input produced by the MCNP code of the RMB core. We investigate the neutron flux values considering a state-of-art high-resolution diffractometer, and the results are promising since some simulated scenarios present values compatible with high-intensity devices.

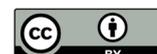

# 1. INTRODUCTION

The Brazilian Multipurpose Reactor (RMB) represents a milestone in the history of Brazilian science. Currently, RMB is the largest and most important drag project for the development of nuclear technology in the country, under the responsibility and coordination of the National Nuclear Energy Commission (CNEN). The main objectives of RMB and associated facilities are the production of radioisotopes and radiopharmaceuticals, aiming to meet the entire national demand of the Unified Health System (SUS) and the private network, irradiation and testing of nuclear fuels and structural materials, and the development of scientific and technological research using neutron beams [1, 2]. RMB will house the National Neutron Laboratory (LNN), the unit responsible for research activities with a suite of 17 instruments that manipulate neutron beams to investigate the structure of various materials [3]. Among the instruments that make up the LNN is the high-intensity neutron diffractometer, called the Flautim (*Schiffornis virescens*). Beside the high-intensity diffractometer Flautim, Araponga (*Procnias nudicollis*) will be on thermal guide one (TG1) of LNN. Powder diffraction is one of the most widely used techniques to study the structural and nanostructural properties of materials. The technique allows determination of long-range structure in polycrystalline materials, short-range atomic structure in disordered or amorphous materials, structural distortions, and any strain and crystal size induced changes to the structure [4]. Neutron powder diffraction has many attractive features, such as neutron penetrative ability, light element sensitivity, isotope dependent scattering, and its magnetic interaction [5]. Resolution and intensity are two essential characteristics of a pow- der diffraction instrument. While the resolution dictates the ability to discern the real space features of the material, the intensity dictates how quickly one can achieve such a measurement, and both are inversely related [6]. High intensity diffractometers require closer approximation to the neutron source, and therefore should be placed as close as possible to the Reactor Face. This is the first constraint that we must fulfill in simulations. Araponga is inspired by the Echidna instrument from the Australian Reactor OPAL-ANSTO [7].

Analogous to OPAL, the RMB project is developed by the Argentinian scientific company INVAP and according to the agreement signed between both parts, the Argentine company should provide, among other requirements, a reactor core configuration that guarantees minimal thermal neutron flux values at the neutron instrument hall of an order of $10^9$ n/cm$^2$s. To prove that the current project fulfills the minimal flux value requirement, INVAP has provided documents and correspondent simulation files. In this sense, the RMB scientific group has received an MCNP input file and its PTRAC (Particle Track Output) file, which corresponds to the neutron flux at the entrance of the thermal guide (TG2). Additional scenarios, such as the neutron flux in other positions of the reflector tank, such as at the entrance of TG1 and TG3, are under the responsibility of the scientific team of the RMB. Since the INVAP MCNP file possesses only commercial purposes, the RMB group must reach Monte Carlo computational autonomy to perform MCNP simulations to design and project neutronic instruments to be built. Such achievement will also allow the RMB team to obtain the performance of every instrument for different core configurations (begin and end of the cycle, control rods configuration, fuel burnup, etc.). The technical development to run these simulations and to get this autonomy is in progress.

In this context, open-source software McStas, which can reed PTRAC files as input, has been widely used to model neutron instruments and guides [8, 9]. The objective of this work is to simulate a simple configuration for Araponga, starting from the basic Echidna configurations. We determine the neutron flux at the sample location to ensure optimal values for data acquisition, and discuss how improvements can be made in order to optical components that can optimize the diffraction technique. With these results, we also intend to investigate if the configuration of instruments among thermal guides, which is given in the basic RMB project, guarantees the best performance for all of them.

## 2. MODELLING ARAPONGA

The linear dimensions of the neutron guides, as well as the relative distance to the Reactor Core and the components that make up Araponga can be found in Figure 1. It is important to note that the Araponga instrument is intended to be installed at the end of the TG1 thermal guide [3]. According to the project documentation, the neutron distribution at the entrance of TG2 possesses a same order flux value, i.e., $10^9$ n/cm$^2$s, and representative neutron divergence and wavelength distribution to those of the TG1 and TG3. In this sense, the PTRAC TG2 file can be used in McStas simulations of the TG1 instrument Araponga as a first approach to determining its components according to the basic engineering project. Since the RMB scientific group is currently working on getting PTRAC for different core moments of fuel cycles and guide entrance positions, it will be possible to get a proper design for all instruments of the basic project. From the results presented in this work, it will be possible to check if the instrument configurations foreseen in the RMB project corresponds to the finest ones considering diffractometer performances. In other words, we could ensure which guide each instrument should be allocated.

**Figure 1:** Scheme of the guide system from the reactor core to the Araponga instrument.

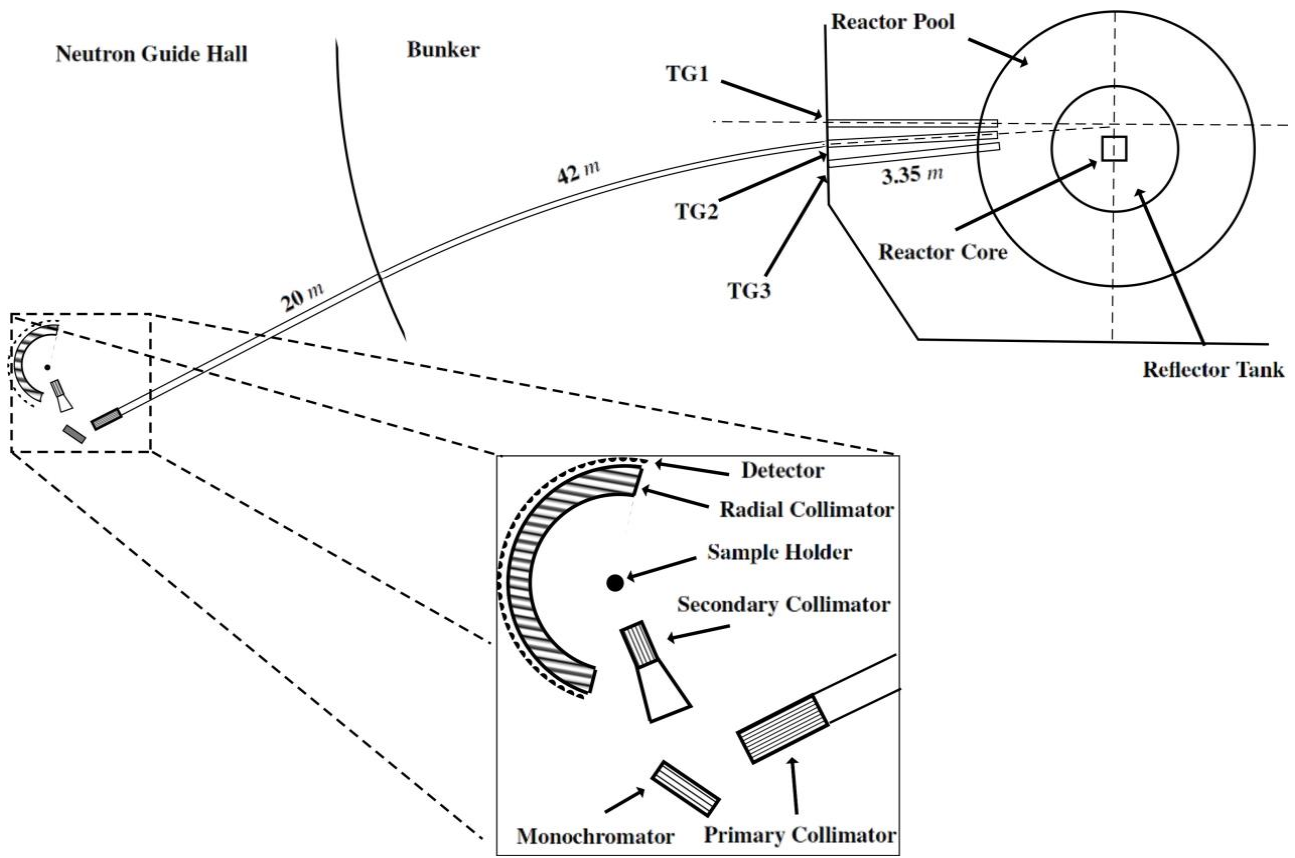

In this work, we use neutron guides in the Bunker with three different curvature values, which were previously calculated to avoid the undesirable direct line of sight (LoS), that is, to ensure that each neutron undergoes at least one reflection on the guide wall [10]. In these simulations, the values of the curvature have been based on two different approaches. The first considers the LoS of the barely curved guide, which is exclusively inside the Bunker area as shown in the scheme of Figure 1. In this case, the curvature that guarantees the LoS is 4410 m. On the other hand, by considering the section of straight guide inside the in-pile portion of the system, we obtain the curvature of 5080 m.

**Table 1:** Components and parameters of McStas simulations

| Components | Parameters and characteristics |
|---|---|
| In-Pile Guide | Section (width x height) = 5 cm x 30 cm; Length = 3.35 m; Curvature = infinity (straight guide); Supermirror index M = 3 (all sides); |
| Bunker Guide | Section (width x height) = 5 cm x 30 cm; Length = 3.35 m; Curvatures = 4410 m, 4745 m, 5080 m; Supermirror indexes: $M_{out}$ = 3, $M_{top}$ = 3, $M_{bottom}$ = 3, $M_{in}$ = 2.5; |
| Hall Guide | Section (width x height) = 5 cm x 30 cm; Length = 20 m; Curvature = infinity (straight guide); Supermirror index M = 2.5 (all sides); |
| Primary Collimator | Section (width x height) = 5 cm x 30 cm; Length = 0.7 m; Divergence = 10'; |
| Monochromator | Section (width x height) = 5 cm x 30 cm; Germanium Monochromator = Vertical Focusing; Crystallographic planes: 0.863 Å (533), 1.089 Å (511), 1.714 Å (311); Take-off angle = 45 degrees; |
| Funnel Guide | Input Section (width x height) = 5 cm x 30 cm; Output Section (width x height) = 5 cm x 13 cm; Length = 1.8 m; Supermirror index M = 2.5 (all sides); |
| Secondary Collimator | Section (width x height) = 5 cm x 13 cm; Length = 0.3 m; Divergence = 10'; |
| Sample | Section (width x height) = 0.6 cm x 4 cm |

The formal calculation of the LoS of a 42 m long curved guide and a 3.35 m straight guide, which conducts neutron flux from reactor pool to Bunker (see Figure 1), has been done utilizing previous deduced geometrical equations [10]. In addition, the third value of curvature has been determined by simply considering the average value of these two curvatures, i.e., 4745 m. The simulation parameters of each component of the diagram in Figure 1 are shown in Table 1. As previously pointed out, the configuration of Araponga has been based on Echidna, the High-Resolution Diffractometer allocated at the OPAL. Simulations have been performed considering three different monochromators for each curvature. Results are displayed in Table 2.

## 3. RESULTS AND DISCUSSION

The results of the Monte Carlo simulations can be found in Table 2. We can see that the flux at the sample position has excellent intensity, compatible even with the best high-intensity neutron diffraction instruments located in international nuclear facilities [6, 7]. We observed that the maximum value was obtained for the configuration R = 4410 m and d = 1.089 Å. We observe that there is a tendency of curvature of 4410 m to possess higher fluxes at the sample place, which is the opposite behavior if one considers just the fluxes that pass through a curved guide. It is worth noting that the monochromator used in simulations corresponds to a vertical focusing model. Thus it is normal to find higher neutron fluxes at the sample place than just after the monochromator. This behavior can be seen in Table 3, which shows fluxes of simulated cases in different parts of the guide system, i.e., at the Neutron Guide Hall (NGH), before and after the primary collimator, after the primary collimator, and at the sample place. Besides, considering that each monochromator d-spacing provides a different selected neutron wavelength. Therefore, one can only compare flux values of the same monochromator. In this sense, it is possible to observe that the curvature value growth imposes higher flux values, except for curvature R = 5080 m and d = 1.089 Å case.

**Table 2:** Monte Carlo simulation results for neutron flux at the sample location for parameter sets. The hkl lattice parameters are the Miller indices.

| R (m) | d (Å) (hkl) | λ (Å) | Flux (×10⁶ n/cm²s) |
|---|---|---|---|
| 4410 | 0.863 (533) | 1.22 | 1.0724 |
|  | 1.089 (511) | 1.54 | 11.305 |
|  | 1.714 (311) | 2.42 | 4.8604 |
| 4745 | 0.863 (533) | 1.22 | 4.1578 |
|  | 1.089 (511) | 1.54 | 3.6852 |
|  | 1.714 (311) | 2.42 | 2.7607 |
| 5080 | 0.863 (533) | 1.22 | 3.0410 |
|  | 1.089 (511) | 1.54 | 5.1145 |
|  | 1.714 (311) | 2.42 | 2.4780 |

**Figure 2:** Neutron flux profile just after the primary collimator for three different curvatures. The black continuous line stands for the curvature of 5080 m, the red dashed line for 4745 m, and the blue dotted line for 4410 m.

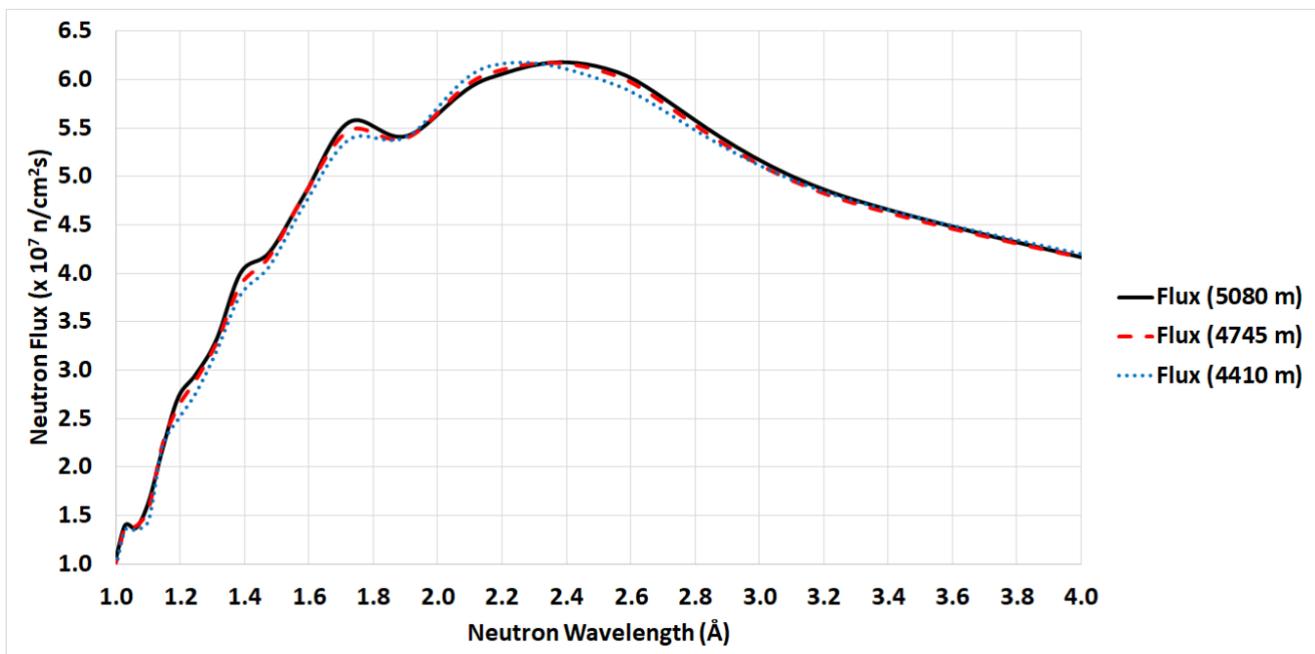

From the results of Table 3, we observe that the curvature of guides does not contribute significantly to higher fluxes since values just at the NGH differ about 2.5% among them, which is in the simulation error. In this sense, the components of instruments, such as collimators, monochromators, etc., are more important to define instrument performance than the guide curvature itself. The difference of fluxes presented by the guide curvatures can be also seen in Figure 2, which represents the neutron flux profile at the NGH, just before the primary collimator.

However, since guide curvatures determine different wavelength neutron transport efficiency and output flux divergence, a direct correspondence between flux values and curvature is not always possible. From that matter, it is important to point out that fine adjustments in the primary and secondary collimators can bring optimal conditions for the diffraction technique. All these aspects are also directly related to the flux spectra that reach the instrument primary collimator. In this scenario, we can see that there is a tendency for neutrons with higher wavelengths to behave similarly when transported by a curved guide. This can be observed in Figure 2 for neutrons with wavelengths higher than 3 Å, where three lines begin to overlap each other. So, adjustments are more necessary for instruments that needs neutrons with lower wavelengths.

**Table 3:** Monte Carlo simulation results for neutron flux at different locations of guide systems, at the NGH, after the primary collimator (Coll1), after the monochromator (Monochr) and at the sample place. The hkl lattice parameters are the Miller indices.

| R (m) | d (Å) (hkl) | Flux NGH ($\times 10^9$ n/cm$^2$s) | Flux Coll1 ($\times 10^8$ n/cm$^2$s) | Flux Monochr ($\times 10^6$ n/cm$^2$s) | Flux Sample ($\times 10^6$ n/cm$^2$s) |
|---|---|---|---|---|---|
| 4410 | 0.863 (533) | 2.4482 | 6.1570 | 3.4484 | 1.0724 |
|  | 1.089 (511) |  |  | 3.8758 | 11.305 |
|  | 1.714 (311) |  |  | 5.2102 | 4.8604 |
| 4745 | 0.863 (533) | 2.4653 | 6.2525 | 3.7640 | 4.1578 |
|  | 1.089 (511) |  |  | 3.7974 | 3.6852 |
|  | 1.714 (311) |  |  | 5.2666 | 2.7607 |
| 5080 | 0.863 (533) | 2.4784 | 6.3111 | 3.9543 | 3.0410 |
|  | 1.089 (511) |  |  | 3.6243 | 5.1145 |
|  | 1.714 (311) |  |  | 5.2964 | 2.4780 |

## 4. CONCLUSION

The simulation results for the neutron flux at the sample site are extremely encouraging. We started with a basic configuration for the Araponga instrument, and many adjustments could be made to optimize the neutron diffraction technique. The results obtained are compatible with state-of-the-art instruments from international installations [6, 7]. Our model will receive new attributes after the simulations for the TG1 thermal guide, where the high intensity neutron diffractometer, called Flautim, will be localized [3]. The optimization problem has many variables, such as the choice of collimators and evaluation of beam divergences in curved sections. Soon, we will present more complete discussions about the wavelength ranges for each studied configuration.

## ACKNOWLEDGMENT

The authors are thankful to the technical coordinator of the Brazilian Multipurpose Reactor (RMB) Project, Dr. J.A. Perrotta. A.P.S. Souza and L.P. de Oliveira also would like to thank CNPq for financial support under grant numbers 381565/2018-1 and 380183/2019-6, respectively.